\documentclass[twocolumn]{article}
\usepackage[english]{babel}
\usepackage[letterpaper,margin=1in]{geometry}

\usepackage{graphicx}
\usepackage{siunitx}
\usepackage[colorlinks=true, allcolors=blue]{hyperref}
\usepackage{color}
\usepackage{amsmath}
\usepackage{nomencl}
\usepackage{array}
\usepackage{hyperref}
\usepackage{authblk}
\usepackage{silence}

\usepackage{amsmath}
\usepackage{nomencl}
\usepackage{authblk}
\usepackage{cite}
\usepackage{comment}

\title{Deployable 3D mesoscale structures through wafer fabrication, geometric frustration and bistable auxeticity}

\author[1]{Yue Wang}
\author[2]{Kelvin Shum}
\author[3]{Yuyang Song}
\author[1,*]{Tian Chen}

\affil[1]{Mechanical and Aerospace Engineering, University of Houston, Houston, TX 77204}
\affil[2]{Mechanical Engineering, University of Texas at Austin, Austin, TX, 78712}
\affil[3]{Toyota Research Institute of North America, Ann Arbor, MI, 48105}
\affil[*]{Corresponding author. Email: tianchen@uh.edu}
\date{}
\begin{document}
\maketitle

\textit{
Transforming flat mesoscale devices into precise three‑dimensional architectures is immensely important for future flexible electronics, biomedical implants, and adaptive optics. However, current methods are limited in fabricating functional free-standing 3D structures using wafer fabrication technologies. We demonstrate this by fabricating polyimide-based architected 2D precursors that, through the bistable nature of its constituent units, deploy into free-standing 3D structures. To impose specific target Gaussian curvature through deployment, we digitally flatten the target mesh using conform maps from 3D to 2D, and locally tune the microstructure of each unit cell such that their respective second equilibria occur at the required conformal scaling factor. This is made feasible by first computing a library of parametric unit cells that that exhibit tunable isotropic expansion and intrinsic bistability. 
The resulting 2D precursor features a spatially heterogeneous tessellation, and can uniquely deploy to the target 3D shapes stably. 
This generative method is first applied to a dome as a pedagogical example. The dome is fabricated as a flat disk and deployed through indentation to 3D. Both deployment accuracy and structural stability are numerically predicted and experimentally validated. During deployment, the structure stays in the predicted 3D shape after the removal of the indenter. By reversing the indenter, we show that the dome recovers to its 3D shape after being pressed at its apex. The same fabrication method is applied to a host of complex 3D shapes with both positive and negative curvatures, and different topologies to showcase versatility. 
Lastly, paraboloidal reflectors with tunable focal lengths are fabricated. Using a collimated laser ray, we demonstrate the resulting reflected patterns agree with geometrical optics. This wafer fabrication-based technology advances beyond flexible electronics and paves the way towards \textit{deployable electronics} to broaden the spectrum of realizable 3D devices.}

\vspace{2mm}


\vspace{2mm}

Three-dimensional mesoscale structures with doubly-curved geometries promise to transform medical engineering, curved electronics, and wireless communication~\cite{xia1998soft,rogers2010materials,liu2017labonskin}. Direct microscopic 3D printing, such as two-photon polymerization, offers geometric freedom but are severely limited in the capability to integrate dissimilar materials and components~\cite{edgar2015additive,harinarayana2021two}.
We demonstrate a more generalized approach to circumvent this limitation by utilizing wafer fabrication. We first pattern a two-dimensional spatially heterogeneous precursor designed according to a user-specified 3D shape. This 2D precursor is then stably deployed via a mechanical stimulus into the target 3D structure~\cite{xu2015assembly,choi2019programming,sim2019three}. 
This strategy can be directly integrated within semi-conductor fabrication protocols, yet allows for the realization of truly free-standing 3D device architectures.

\begin{figure}[t!]
	\centering
	\includegraphics[width=82.5mm]{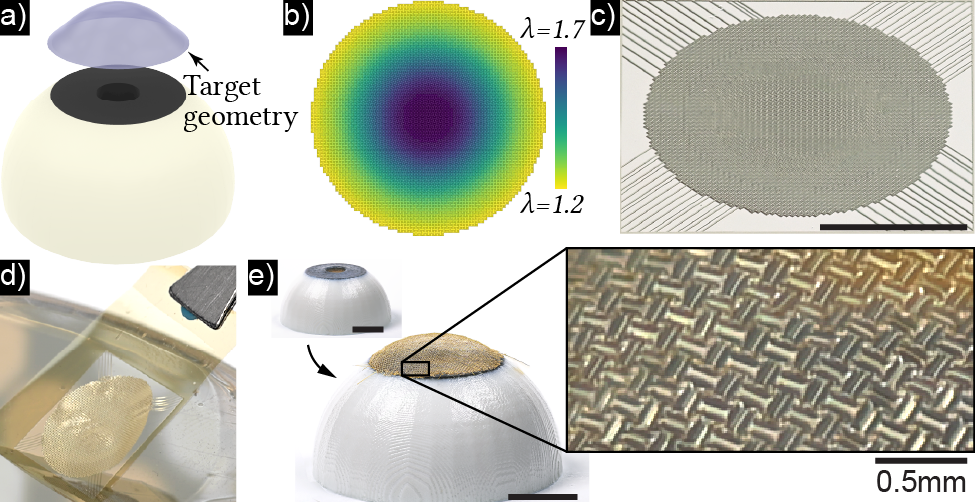}
	\caption{\textbf{Pipeline for the manufacturing of deployable 3D mesoscale structures.} a) A 3D doubly-curved target geometry is specified as a surface mesh. We use the example of a device approximating the geometry of a human cornea. b) The mesh is mapped conformally to the 2D plane. The color spectrum indicate the necessary isotropic expansion $\lambda$ per mesh face. A grid of square units with auxetic ($\nu=-1$) and bistable microstructure is tiled onto this map. Each unit cell is geometrically tuned to exhibit isotropic expansion upto a local strain energy minium indicated by the equilibrium strain $\varepsilon_\mathrm{equil}=\lambda-1$. c) A multi-step lithography process is employed to fabricate the 2D precursor. d) This is physically lifted from the wafer substrate and e) deployed into the prescribed 3D shape. Scale bars are all 5mm unless indicated otherwise.}\label{fig:1}
\end{figure}

To achieve this, we address three fundamental challenges related to 1) accommodating the necessary mechanical stretch, 2) targeting specific 3D shapes, and 3) ensuring structural stability. First, Gauss's Remarkable Theorem dictates that in-plane stretch and/or contraction is required to alter the Gaussian curvature of a surface~\cite{hicks1965notes}. To achieve significant curvature change, the necessary stretch often exceeds the rupture limits of thin film substrates~\cite{evans1991molecular,pressley2010gauss}. To overcome this, one may constrain the designs to isometric deformations or bending ~\cite{xu2015assembly,guo2018controlled,dong2023microfabrication,liu2024threedimensionally}. Alternatively, a microstructural approach such as Kirigami, serpentine, or origami patterns may be adopted to exhibit a larger effective stretch of the overall material~\cite{shenoy2012self,blees2015graphene,lamoureux2015dynamic,xu2017highly,jin2024engineering,felton2015selffolding}.

Second, current techniques generally do not allow target geometries to be prescribed \textit{a priori}. By virtue of periodic patterning, most 2D precursor designs are inherently 3D-shape-agnostic, \textit{i.e.}, they may equally conform to many different substrate geometries. As a corollary, it is also unknown \textit{a priori} whether the required stretch of a specific contour exceeds the allowable stretch of the precursor, thereby causing tear and/or wrinkling~\cite{xu2015assembly,konakovic2016developable,guo2018controlled}.
Instead, we construct a conformal map of the target 3D shape and generate a spatially varying 2D pattern accordinate. This 2D pattern can only deploy to that specific 3D shape up to isometry.

\begin{figure*}[t!]
	\centering
	\includegraphics[width=165mm]{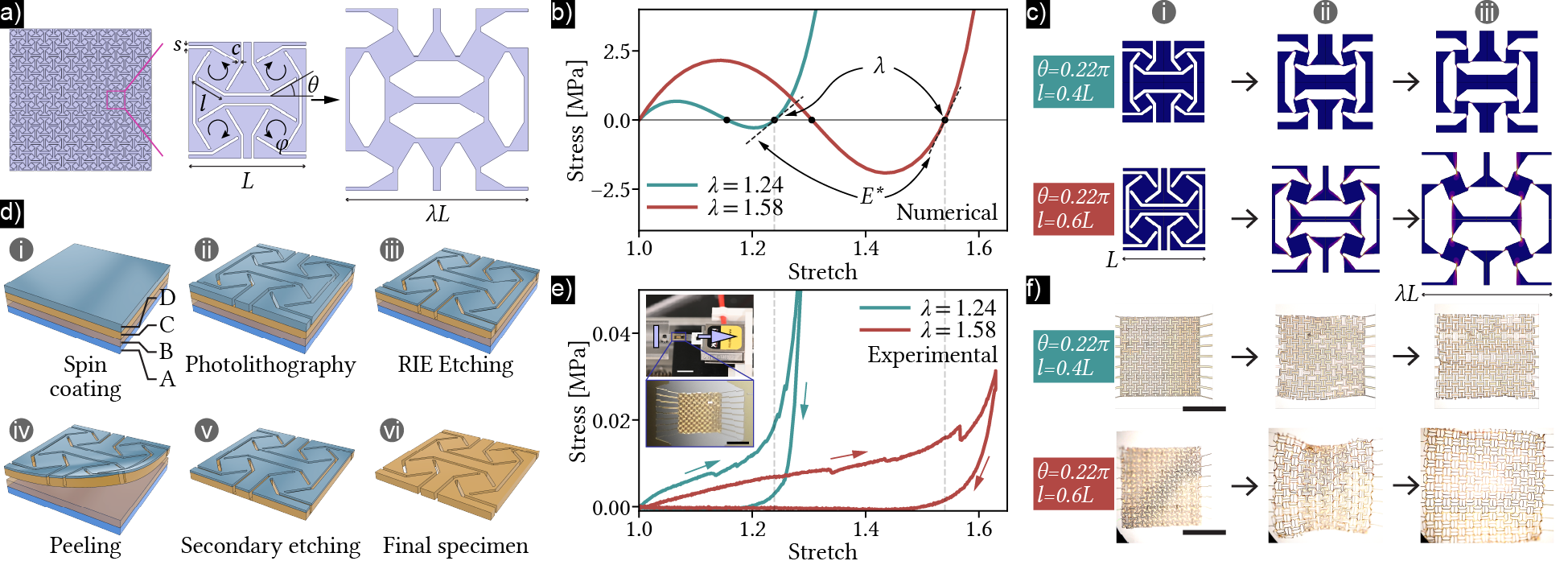}
	\caption{\textbf{Mechanics of bistable auxetic unit cells.} a) Each 2D pattern is tessellated using a grid of unit cells. The microstructure of each unit cell features a number of slits. As the unit cells stretch, the inner squares counter-rotate. By changing internal geometric parameters $l$ and $\theta$, the stretch at the second equilibrium $\lambda$ can be tuned. b) From all combinations of $l$ and $\theta$, we select two extremes of the $\lambda$ spectrum.
    Finite element (FE) predicted stress-stretch curve of these unit cells show that the stretch at the second equilibrium can be tuned. c) The corresponding deformation as a result of FE simulations uses periodic boundary conditions. d) Wafer fabrication protocol used by all specimens in this work. e) Experimental stress-stretch curves showing 8 by 8 tessellations under tension. Scale bar is 1 mm. f) Microscopic image sequence of the deformation.}\label{fig:2}
\end{figure*} 

Third, the mechanical stability of flexible electronics in the absence of a substrate remains undefined~\cite{py2007capillary, ko2009curvilinear,sim2019three, cheng2023programming}, \textit{e.g.}, when the substrate is removed, most flexible electronics may collapse back to planar configurations.
By engineering a second stable state in the target 3D shape, we can define the mechanics of the 3D mesoscale structures after shape deployment~\cite{haghpanah2016multistable,rafsanjani2016bistable,chen2021bistable}. This advancement will enable self-equilibrated, load-bearing micro-architectures, such as miniature antennas, conformal sensor arrays, and micro-optical components~\cite{ko2009curvilinear, rao2021curvy,cheng2023programming}.

Using a cornea transplant geometry as an example, we demonstrate a protocol that utilizes wafer fabrication to manufacture a 2D precursor, which is then deployed into a free-standing doubly curved 3D structure. The protocol first digitally model a 3D mesh of the target cornea geometry. 
A discrete conformal map is then computed~\cite{sawhney2017boundary} to result in a 2D mesh with the identical connectivity as the 3D mesh. The vertices, however, are mapped to the 2D plane and optimized to minimize shape distortion of the triangular mesh faces, \textit{i.e.}, the internal angles of each triangle are as preserved as possible. This necessarily requires the total area of the triangles to isotropically scale, as visualized in Fig.~\ref{fig:1}b where the expansion factor $\lambda$ at the center is larger than those at the circumference. This areal scaling is mechanically enabled using a parametric unit cell that exhibits auxeticity and bistability with a tunable second equilibrium strain, $\varepsilon_\mathrm{equil}$. By setting this strain to $\varepsilon_\mathrm{equil}=\lambda-1$, we achieve the necessary scaling per unit cell. Through a multi-step lithography process using polyimide PI 2611 (HD Microsystems Inc.), we manufacture this 2D precursor and lift it off the substrate (Fig.~\ref{fig:1}d). This 2D precursor is then mechanically deployed into the 3D target shape. To demonstrate its stability in 3D, we place the deployed structure upon a 3D printed eye model without the cornea, \textit{i.e.}, the flat iris is exposed (Fig.~\ref{fig:1}e).

First, we analyze the kinematics and mechanics of the unit cell microstructure that enables auxeticity and bistability. 
All deployable surfaces in this work are tessellated with unit cells of the same initial edge length $L\approx \SI{220}{\micro\metre}$ (Fig.~\ref{fig:2}a). A series of slits within each unit cell as defined by the side length of $l$ and  a tilting angle $\theta$. The slits have a width of $s$, and they form four inner squares connected to the outer geometries with hinges of width $c$. The detailed geometric definition is found in SI Sec.2.1.
Under uniaxial tension, these squares counter-rotate by an angle $\phi$ from $0$ to $2\theta$ to accommodate the stretch. 
Kinematically, this results in an isotropic expansion of the unit cell from the initial edge length $L$ to $\lambda L$~\cite{chen2021bistable} where $\lambda_\mathrm{kinematic}=1+2 l \sin{\theta}$ assuming $c,s=0$. 

Note that at all intermediate values of $\phi$, kinematic incompatibility must be resolved through deformation.
We use Finite Element simulations (Abaqus) to provide a predictive model of the mechanical behavior of the unit cell by sweeping combinations of parameters $\theta$ from $0.1\pi$ to $0.28\pi$ and $l$ from $0.1L$ to $0.4L / (\sin\theta + \cos\theta)$ with increments of $0.01\pi$ and $0.01L$ respectively. Each unit cell microstructure is meshed using quadratic shell elements. Periodic boundary condition is imposed on the four edges, and a stretch is prescribed from $1$ to $1.2\lambda_\mathrm{kinematic}$. The material behavior of polyimide is detailed in SI Sec.1. The detailed Finite Element (FE) protocol is outlined in Materials and Methods and SI Sec.3.1.

The resulting stress-strain relationships inform the bistability of the different unit cells as well as the stretch at their second equilibria $\lambda$ (if any). 
Of the parametric combinations of $\theta$ and $l$ that result in a bistable microstructure as defined by the presence of a second local minimum in the strain energy, we identify the permissible spectrum of $\lambda$ as between $[1.2,1.6]$. The stress-stretch behavior of the two ends of this spectrum are shown in Fig.~\ref{fig:2}b. The corresponding deformation exhibit isotropic expansion up to a prescribed $\lambda L$ (Fig.~\ref{fig:2}c). The stress at the hinges are measured to ensure no significant plasticity occurs. See SI Sec.3.2 for the results of all parametric combinations. 

To facilitate experiment characterization at the micron-scale, we leverage wafer fabrication to manufacture all specimens in this manuscript including 8 by 8 tessellations of the different parametric variations. As shown in Fig.~\ref{fig:2}d, first, (i) the wafer is coated with (A) a silane coupling agent (3-Aminopropyl)triethoxysilane (APTES), (B,C) two layers of polyimide (PI-2611, \SI{6}{\micro\metre}), and (D) a photoresist (AZ 12XT-20PL, \SI{11}{\micro\metre}), (ii) then maskless photolithography is used to pattern the microstructure onto the photoresist, (iii) the sample is then uniformly etched until the depth of the slit on the top layer of PI is reduced by 75\% using Reactive-ion etching (RIE). It is not etched through to minimize stress concentration during the peeling process that follows. In (iv), the upper layer of PI is mechanically peeled from the bottom, and (v) subjected to secondary etching until the slits are fully formed. (vi) Lastly, the sample is immersed in acetone to dissolve the remaining photoresist. This protocol is designed to minimize stress concentrations that occur in microscopic fabrication of architected materials when they must detach from substrates. Refer to SI Sec.4 for detailed fabrication protocol.

\begin{figure*}[t!]
	\includegraphics[width=165mm]{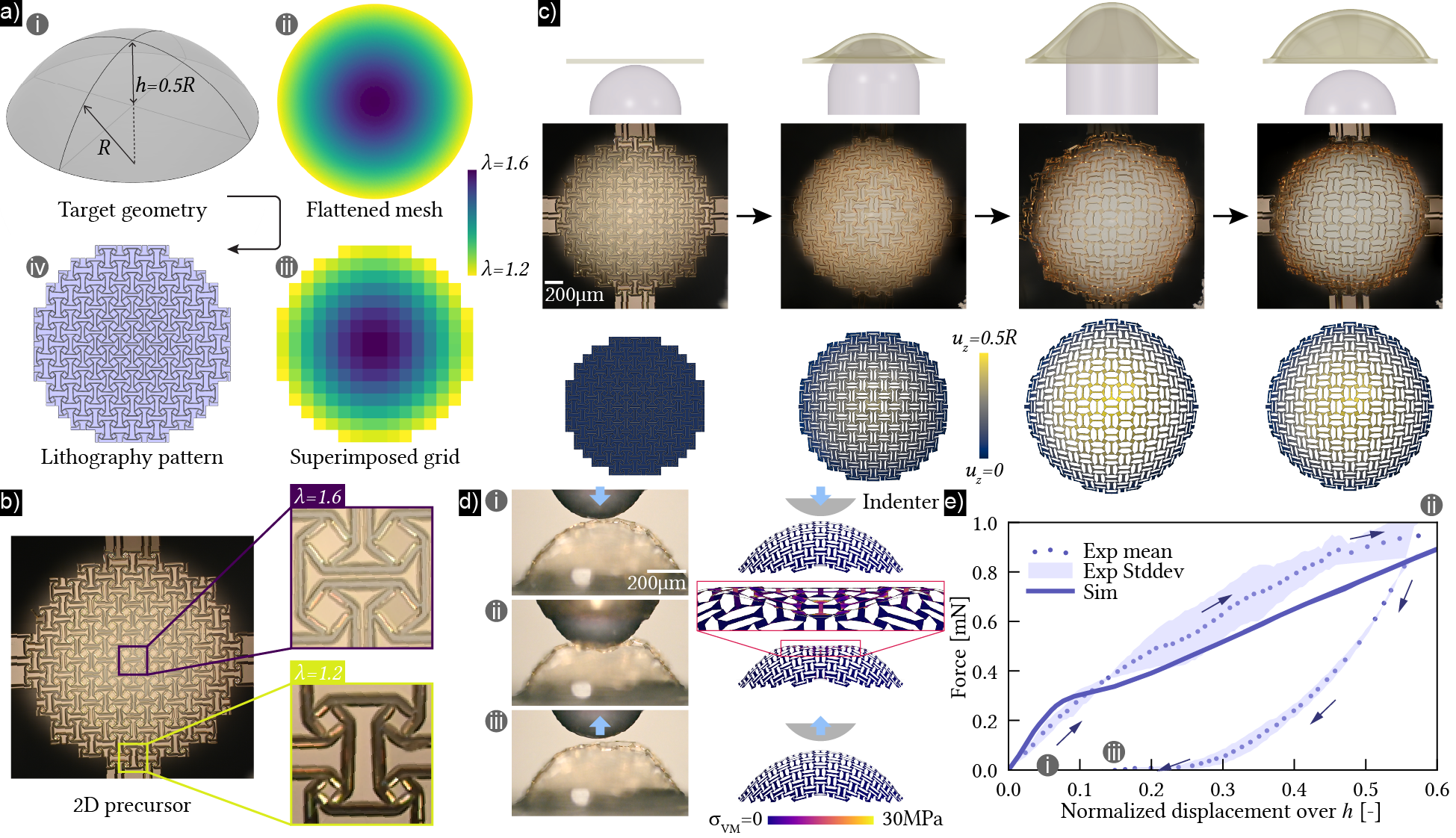}
	\caption{\textbf{Design, fabrication and mechanical behavior of a deployable 3D quarter spherical dome.} a) Digital geometric processing that translates a target 3D mesh to a lithography mask by first conformal mapping to the 2D plane, grid sampling the scale factor $\lambda$, and then selecting the geometric parameters per unit cell. b) Wafer fabrication of the 2D spatially heterogeneous precursor showing unit cells of different $\lambda$. c) 2D to 3D deployment using a mechanical indenter. This is a part of fabrication process to result in a 3D structure. d) Mechanical indentation test to demonstrate the stability of the 3D structure after deployment. e) Experimental and simulation results of the indentation test. See SI Video 2}\label{fig:3}
\end{figure*}

A micro mechanical stage is constructed to apply tension to the 8 by 8 tessellations (See SI Sec.5.1). The experiment is conducted under an optical microscope to measure the effective stretch. The boundary cells are connected with angled beams (Fig.~\ref{fig:2}e inset) which are designed to allow free expansion in the orthogonal direction, \textit{i.e.}, at the second equilibrium, the beams will be parallel. The same two microstructures are experimentally tensioned to $1.2$ times their respective second equilibria and relaxed. The resulting effective-stress and stretch curves show an initial raise followed by a plateau. The initial behavior is attributed to the partial opening of all unit cells as expected from linear elasticity. The plateau is a result of the column by column triggering of bistability (Fig.~\ref{fig:2}f, see Video 1). Due to the imperfections in fabrication and the boundary conditions, the triggering force per column differ minutely. This causes some columns of unit cells to snap prior to others. At the expected $\lambda$, all unit cells are in their second equilibria, and further tensioning results in a significantly stiffer response. During unloading past the second equilibria, the reaction force 
reaches zero as indicative of static equilibrium. With the parametric sweep, we have constructed a library of unit cells that feature the correspondence between the stretch at the second equilibrium $\lambda$ and the input geometric parameters $\theta$ and $l$. Further, we have identified the range of $\lambda$ values that will result physically realizable microstructures that exhibit bistability. 



Now, we discuss an algorithmic approach to design a spatially heterogeneous 2D tessellation that deploys to the prescribed 3D shape when all of its unit cells reach their respective second equilibrium. We observe that all unit cells in the design space expand isotropically up to their respective second equilibrium with a stretch equaling to $\lambda$. This leads us to use conformal mapping to calculate the necessary scalar field of $\lambda$ as a function of the prescribed 3D shape. Once $\lambda$ is calculated per coordinate on the 2D mesh, the values are then further translated to the unit cell geometric parameters $\theta$ and $l$ using the unit cell library computed previously. 

A conformal map is a function that locally preserves angles but not areas. Mechanically, this implies each material point may scale in area but not distort in shape, which our unit cells satisfy. Thereby, if we abstract the prescribed 3D shape by its Gaussian curvature $K$, we can simply relate $K$ directly to the Laplace of the scalar field $\lambda$ (Eq.~\ref{eq:K}),
\begin{equation}
    K=\Delta_f \log{\lambda},
    \label{eq:K}
\end{equation} 
where, $\Delta_f$ is the Laplace-Beltrami operator~\cite{ben-chen2008conformal,konakovic2016developable}.

Mechanistically speaking, the spatially varying degree of stretch on a surface will cause geometric frustration that ``pop'' the surface out into 3D~\cite{dudek2025shape}. This occurs when out-of-plane bending becomes more energetically favorable as compared to in-plane compression of the constituent material.
Indeed, shape morphing that explicitly leverage this principle have been demonstrated~\cite{nath2003genetic,klein2007shaping,zhang2025geometrically}. 

To outline the algorithm, we target the deployment from a disk to a quarter spherical dome, $h=0.5R, R=\SI{1.35}{\milli\metre}$ (Fig.~\ref{fig:3}a,i). First, we algorithmically generate the 2D precursor, which features a spatially heterogeneous unit cell tessellation.
While stereographic projection may be used to calculate a continuous $\lambda$ field for such a canonical geometry~\cite{whittaker1984stereographic}, we use Boundary First Flattening (BFF)~\cite{sawhney2017boundary} for generality in processing arbitrary discrete geometries. BFF computes conformal maps of input meshes by minimizing angular distortion while flattening from 3D to 2D. The scaling $\lambda^2$ is then the areal change of each pair of corresponding mesh faces (Fig.~\ref{fig:3}a,ii).
Next, a grid of square of edge length $L$ is overlaid on the conformal map. By inferring the $\lambda$ of each square from the mesh faces underneath, we calculate the necessary $\theta$ and $l$ of the unit cell microstructure to slot in the square (Fig.~\ref{fig:3}a,iii,iv). 
See SI Sec.2.2 for the derivation using stereographic projection and SI Sec.2.3 and 2.4 for the generative algorithm.

Should the $\lambda$ resulting from BFF exceed the bounds of the permissible $\lambda$ spectrum from the unit cell analysis, then no combinations of $\theta$ and $l$ can satisfy the required scaling, and the target geometry cannot be deployed to with this particular microstructure without injecting singularities~\cite{konakovic-lukovic2018rapid}.

The wafer fabrication protocol outlined in Fig.~\ref{fig:2}d is used to fabricate all 2D precursor patterns (Fig.~\ref{fig:3}b). For deployment, a spherical indenter ($r_\mathrm{indenter} = \SI{0.4}{\milli\metre}$) is used. The 2D pattern is fixed at the mid point of the four edges. The indenter connected to a linaer translation stage pushes up against the pattern until the vertical displacement reaches $d=1.2h$. This is larger the height of the dome to ensure each unit cell is stretched beyond their respective second equilibrium (Fig.~\ref{fig:3}c). The indenter is then retracted to show that the dome similarly retracts briefly until the intended height is reached, then detaches from the indenter. To show that the 3D structure is not shaped as a result of indentation, we fabricate a periodic tessellation of unit cells following the same 2D outline. Following the identical deployment process, the structure remains 2D except where confined by the boundary points (See SI Sec.5.4).

Using shell elements and static analysis, the FE simulation of deployment artificially expands the distance between every pairs of neighboring unit cells from $L$ to their respective $\lambda L$ where the $\lambda$ of the each pair is averaged. The total strain energy shows two local minima during the simulation, demonstrating the stability of the deployed 3D structure (see SI Sec.3.3). This method of simulated deployment can be universally applied as the target 3D shape is not supplied \textit{a priori}. By not applying displacements at the boundary nodes, this method circumvent the instabilities inherent in each unit cell. We find the resulting deployed 3D structure matches closely to the experimental counterpart. See Materials and Methods for details of the FE setup.

Lastly, to demonstrate structural stability, the indenter is flipped to press downward onto the deployed dome as it rests against a flat surface. The displacement of the indenter is increased until $d=0.6h$ upon which it reverses. Microscopic imaging of the deformation shows a behavior similar to those of continuum shells where a localized indent forms~\cite{taffetani2018static}. Upon reversal of the indenter, the recovery of the dome structure (Fig.~\ref{fig:3}d) demonstrates its structural stability.
The FE counterpart replicates the experimental setup and allows access to the internal stress state of the dome. The force displacement behavior of the simulation matches well with that of the experiment for the loading portion (Fig.~\ref{fig:3}e). As the simulation model does not account for frictional dissipation (of the dome against the flat surface), the unloading is not predicted.



\begin{figure}[t!]
	\centering
	\includegraphics[width=82.5mm]{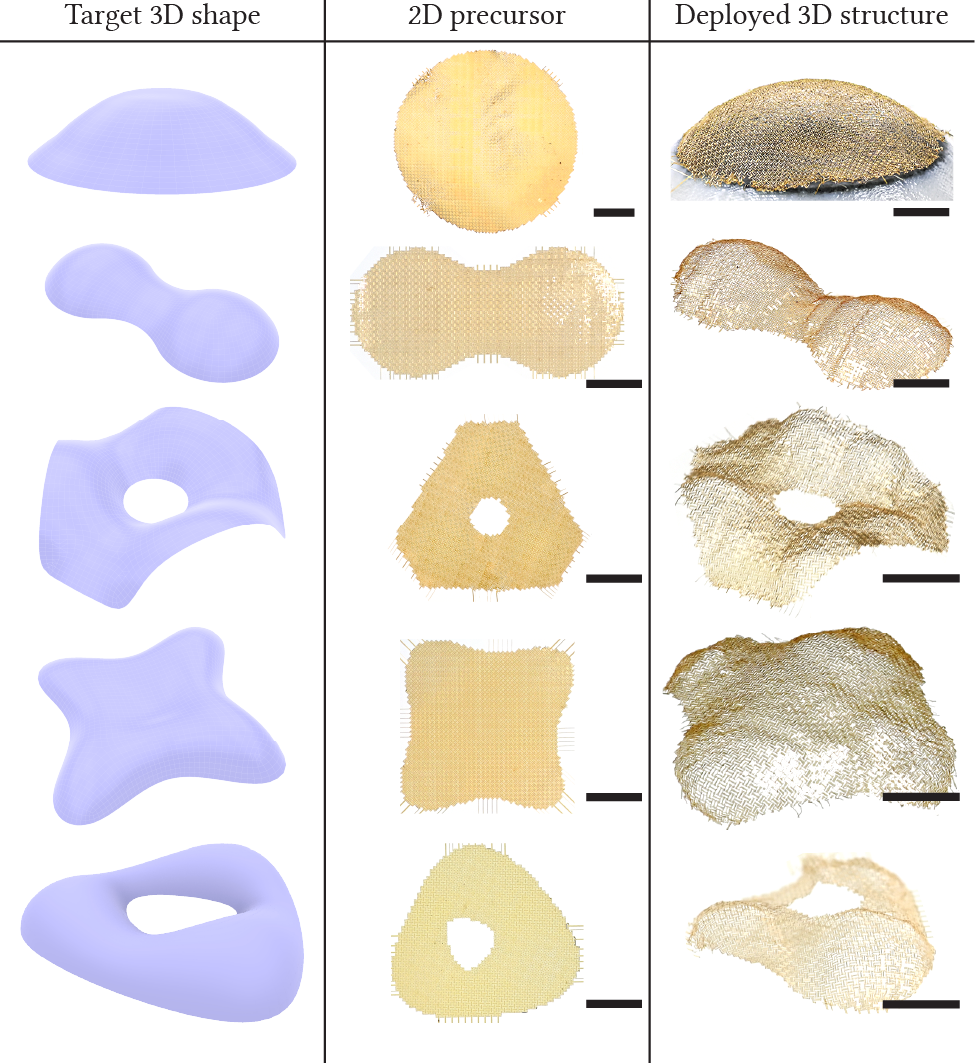}
	\caption{\textbf{Experimental demonstration of complex target 3D shapes.} The 3D meshes are supplied to the algorithm which generates the corresponding 2D precursor lithography patterns. These are fabricated and deployed to 3D structures of prescribed shapes. Scale bars: \SI{2}{\milli\metre}.}\label{fig:4}
\end{figure}

To demonstrate the versatility of the 2D to 3D deployment methodology, we specify a number of target shapes with different topology and geometry (Fig.~\ref{fig:4}). These include the cornea geometry shown in Fig.~\ref{fig:1}, a double dome with a negative curvature region in between~\cite{chen2021bistable}, a 3-fold rotational symmetric shape with open boundaries that are elevated~\cite{chen2021bistable}, the top of the Lilium tower~\cite{fairs2008lilium}, and an annulus~\cite{panetta2021computational}. The Euler characteristic of shapes 3 and 5 are 0 whereas the others are 1~\cite{richardson2014efficient}. In each case, the 2D precursors are fabricated in the 2D state with some boundary units connected to the substrate. The five shapes are designed using 1126, 667, 501, 590, 515 unit cells respectively.
Upon mechanical deployment using indentation, the 3D structures faithfully reproduce the target input geometries (See SI Video 3).

\begin{figure}[ht!]
	\centering
	\includegraphics[width=82.5mm]{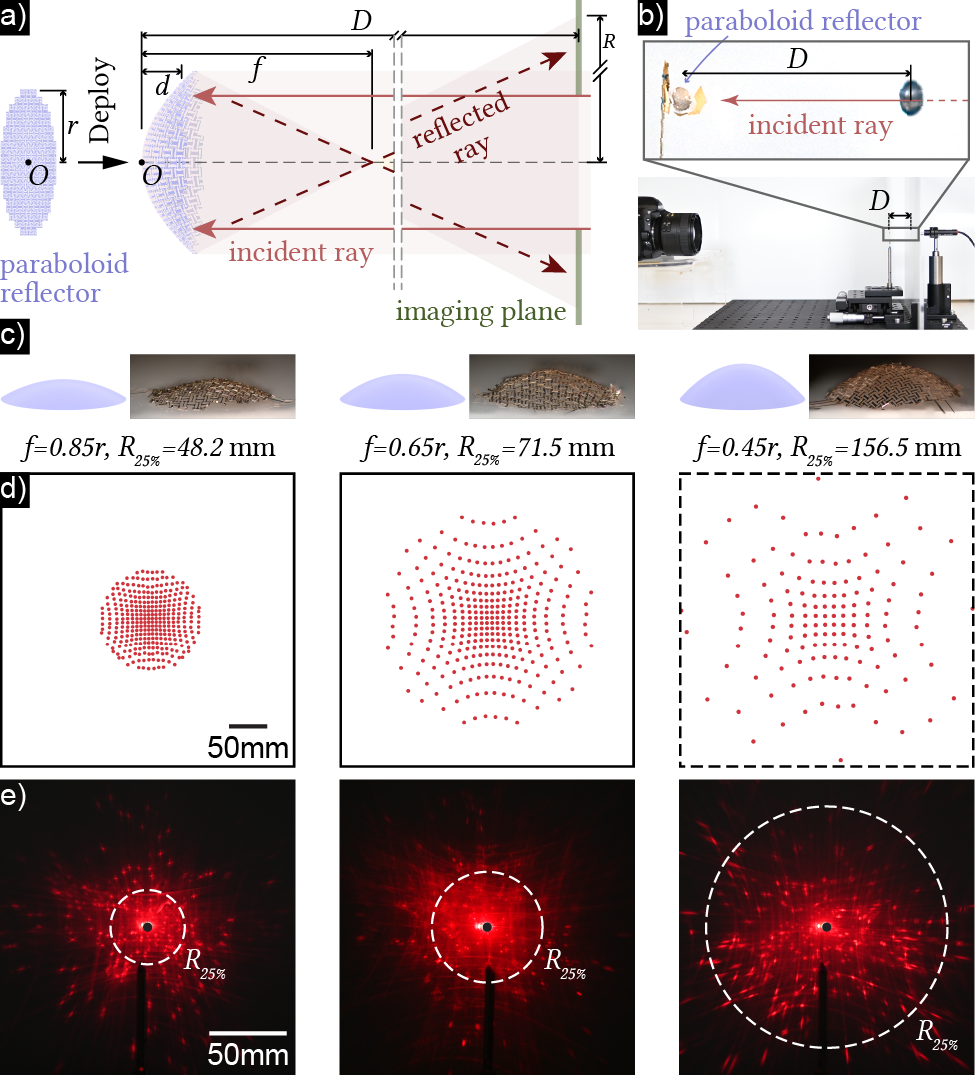}
	\caption{\textbf{Demonstration of proof-of-concept doubly-curved paraboloidal reflectors with tunable focal distances.} a) The schematic diagram of experimental setup of the paraboloidal reflector shows the incident rays reflecting off the deployed reflector, converging at the focal point before diverging towards the imaging plane. b) Real-scale of the experimental setup. c) Target 3D shape, and the fabricated and deployed 3D reflectors with focal lengths ranging from $f=0.85r$, $0.65r$ to $0.45r$. d) Analytically calculated reflected pattern for the three samples at $D=\SI{40}{\milli\metre}$, note that the reflected pattern of specimen $f=0.45r$ exceeds the bounding box. e) Photography of the reflected patterns with the analytical $R_{25\%}$ radii labeled.} \label{fig:5}
\end{figure}

We use the example of paraboloidal reflectors to demonstrate precise control of curvature of the deployed 3D structures. Here, we fabricate reflectors of three different focal lengths $f = 0.85r, 0.65r, 0.45r$ with a constant radius $r=\SI{3}{\milli\metre}$. The aforementioned algorithmic wafer fabrication is adopted. To increase the reflectivity of the translucent polyimide, a thin layer of gold is coated on the 2D precursor as a final step. All three patterns feature the same number of unit cells and the same edge length $L\approx \SI{220}{\micro\metre}$.

The optical experimental schematic shows the arrangement of the deployed 3D paraboloidal reflector (Fig.~\ref{fig:5}a). A visible laser of \SI{3}{\milli\metre} in diameter (Thorlabs, PL202, \SI{635}{\nano\metre}, \SI{1}{\milli\watt}) illuminates the entire surface of the reflector with collimated rays. The rays are reflected and captured on the opposing imaging plane. The reflector is positioned $D=\SI{40}{\milli\metre}$ away from the imaging plane, with a digital camera placed behind (Fig.~\ref{fig:5}b). With the $f$, $r$ and $D$ known, the shape of the reflected pattern can be calculated with geometric optics. The paraboloidal shape will converge all reflected rays at the target focal length before they diverge and intersect with the imaging plane.

By tracing all the rays that reflect off the inner rotating squares of each unit cell, we determine the light pattern on the imaging plane (Fig.~\ref{fig:5}c).
The predicted patterns show an accumulation of reflected rays around the center for all patterns. Away from the center, due to the arrangement of the square unit cell in the deployed state, the reflected rays exhibit a diagonal pattern.
The predicted diameter of the illuminated area for the three focal lengths of $0.85r$, $0.65r$ and $0.45r$ are $R=\SI{69.5}{\milli\metre}$, \SI{147}{\milli\metre}, and \SI{381}{\milli\metre} respectively. Please refer to SI Sec.2.5 for the calculations. Experimentally, the reflected patterns are captured under identical lighting conditions (Fig.~\ref{fig:5}d). The spread of the reflected rays match those of predictions, showing that the deployed 3D structure closely approximated the target paraboloidal shape and curvature.
For cases where \( f = 0.45r \), the radius \( R \) of the illuminated region exceeds the imaging frame. Instead, we define and visualize an effective radius \( R_{25\%} \), which encloses exactly 25\% of the total reflected intensity for each pattern. Note that as the total intensity is invariant, the brightness of $f=0.45r$ is much less than those of the others. 

This study investigates the geometry and mechanics of deployable 3D mesoscale structures. Specifically, we generate optimized lithography patterns to wafer fabricate 2D precursors that precisely deploy to target 3D structures. Going beyond mechanical deployment, future research may investigate actuation mechanisms including magnetism~\cite{kim2018printing} or pneumatic inflation~\cite{panetta2021computational} to create next generation deployable target aware semi-conductor devices~\cite{rogers2010materials}. 

\section*{Materials and Methods}

\paragraph{Fabrication protocol} 
All specimens are fabricated using the protocol outlined in Fig.~\ref{fig:2}d. The ingredients are as follows. The silane coupling agent is (3-Aminopropyl)triethoxysilane (APTES). The structural material is polyimide (PI-2611). The photoresist used is AZ 12XT-20PL. The maskless photolithography system is Bruker SF-100 Lightning. The Reactive-ion etching is an Oxford Plasmalab System 100 ICP. See SI Sec. 4 for details.

\paragraph{Material characterization} The stress-strain behaviors of polyimide is show the a universal testing  machine (Instron 68SC-1). Testing velocity is \SI{5}{\milli\metre\per\second}. The results are shown in SI Sec.1.

\paragraph{Experimental setup} Mechanical tests are performed using a custom-built meso-scale testing system. The setup includes a linear motor stage (Zaber) and a force sensor (LSB200, 10g, Futek), mounted onto the stage via a custom-designed 3D-printed adapter. Deformation is observed using a Nikon microscope equipped with a digital camera (Nikon D780). The entire setup is mobile and can be transported to different imaging equipment. A detailed description of the testing procedure and equipment configuration is provided in the SI Sec.5.

\paragraph{Numerical Simulations} 
All Finite Element simulations are conducted using Abaqus static (general), accounting for geometric non-linearity. The polyimide material is modeled as linear elastic. Quadratic shell elements are used to mesh all geometries in this study. The unit cell characterization is modeled with periodic boundary conditions imposed on the four edges. The deployment simulation involves establishing axial connectors between every neighboring pairs of unit cells. Displacements are then assigned to each connector to locally expand each unit cell in plane. The magnitude of the displacement is calculated by the pre-calculated second equilibrium strain of the unit cells. The indentation simulation is a continuation of the deployment, and models a stiff sphere pressing down against the deployed geometry. See SI Sec.3 for details.

\bibliographystyle{elsarticle-num} 
\bibliography{ref}

\paragraph{Acknowledgments} 
The authors gratefully acknowledge the resource and guidance of the University of Houston Nanofabrication Facility and the Shared Equipment Authority (SEA) at Rice University.

\paragraph{Funding} Y.W.,  K.S., and T.C. acknowledge the support by Toyota Research Institute of North America and NASA MIRO ``Inflatable Deployable Environments and Adaptive Space Systems'' (IDEAS$^2$) Center under Grant no. 80NSSC24M0178.

\section*{Author Contributions}
\noindent {Conceptualization:} Y.S., T.C.
\noindent {Methodology:} Y.W., T.C.
\noindent {Investigation:} Y.W., K.S., T.C.
\noindent {Visualization:} Y.W., K.S., T.C.
\noindent {Supervision:} Y.S., T.C.
\noindent {Writing-original draft:} Y.W., Y.S., T.C.
\noindent {Writing-review \& editing:} Y.W., Y.S., T.C.

\section*{Competing Interests}
\noindent The authors declare they have no competing interests.

\section*{Data and Materials Availability}
\noindent All data needed to evaluate the conclusions in the paper are present in the paper and/or the Supplementary Materials.

\end{document}